\documentclass[twocolumn,prl,superscriptaddress]{revtex4}
\usepackage{graphics}
\usepackage{graphicx}
\usepackage{amsfonts}
\usepackage{textcomp}
\usepackage{amssymb}
\usepackage{mathrsfs}
\usepackage{amsmath}
\usepackage{color}
\usepackage{float}

\begin{document}

\title{Universal Hyperuniform Organization in Looped Leaf Vein Networks}

\author{Yuan Liu\footnote{These authors contributed equally to this work.}}
\affiliation{Department of Physics, Qufu Normal University, Qufu, Shandong Province, China 273165}

\author{Duyu Chen\footnotemark[1]}
\affiliation{Materials Research Laboratory, University of California, Santa Barbara, California 93106, United States}

\author{Jianxiang Tian}
\email[correspondence sent to: ]{jxtian@qfnu.edu.cn}
\affiliation{Department of Physics, Qufu Normal University, Qufu, Shandong Province, China 273165}

\author{Wenxiang Xu}
\email[correspondence sent to: ]{xwxfat@gmail.com }
\affiliation{Institute of Solid Mechanics, College of Mechanics and Engineering Science, Hohai University, Nanjing 211100, P.R. China}

\author{Yang Jiao}
\email[correspondence sent to: ]{yang.jiao.2@asu.edu}
\affiliation{Materials Science and Engineering, Arizona State
University, Tempe, AZ 85287} \affiliation{Department of Physics,
Arizona State University, Tempe, AZ 85287}


\begin{abstract}
Leaf vein network is a hierarchical vascular system that transports water and nutrients to the leaf cells. The thick primary veins form a branched network, while the secondary veins can develop closed loops forming a well-defined cellular structure. Through extensive analysis of a variety of distinct leaf species, we discover that the apparently disordered cellular structures of the secondary vein networks exhibit a universal hyperuniform organization and possess a hidden order on large scales. Disorder hyperuniform (DHU) systems lack conventional long-range order, yet they completely suppress normalized large-scale density fluctuations like crystals. Specifically, we find that the distributions of the geometric centers associated with the vein network loops possess a vanishing static structure factor in the limit that the wavenumber $k$ goes to 0, i.e., $S(k) \sim k^{\alpha}$, where $\alpha \approx 0.64 \pm 0.021$, providing an example of class III hyperuniformity in biology. This hyperuniform organization leads to superior efficiency of diffusive transport, as evidenced by the much faster convergence of the time-dependent spreadability $\mathcal{S}(t)$ to its long-time asymptotic limit, compared to that of other uncorrelated or correlated disordered but non-hyperuniform organizations. Our results also have implications for the discovery and design of novel disordered network materials with optimal transport properties.

\end{abstract}
\maketitle

Recently, it was discovered that apparently disordered organizations in biological systems across scales, such distributions of photoreceptor cones in avian retina \cite{Ji14}, self-organization of immune cells \cite{Ma15}, and large-scale patterns of vegetation \cite{ge2023hidden}, can possess an exotic hidden order called ``hyperuniformity''. Disorder hyperuniform (DHU) systems lack conventional long-range order, yet they completely suppress normalized infinite-wavelength density fluctuations like crystals \cite{To03, To18a}. In this sense, DHU systems can be considered to possess a ``hidden order'' in between that of a perfect crystal and a completely random system (e.g., an ideal gas). DHU is characterized by a local number variance $\sigma_N^2(R)$ associated with a spherical window of radius $R$ that grows more slowly than
the window volume (e.g., $\sim R^d$ in $d$-dimensional
Euclidean space) in the large-$R$ limit \cite{To03, To18a}, i.e., $\lim_{r\rightarrow \infty} \sigma_N^2(R)/R^d = 0$. This is equivalently manifested as the vanishing static structure factor in the infinite-wavelength (or zero-wavenumber) limit, i.e., 
$\lim_{|{\bf k}|\rightarrow 0}S({\bf k}) = 0$,
where ${\bf k}$ is the wavenumber and $S({\bf k})$ is related to the pair-correlation function $g_2({\bf r})$ via $S({\bf k}) = 1+\rho \int e^{-i{\bf k}\cdot {\bf r}}[g_2({\bf r}) - 1]d{\bf r}$ and $\rho = N/V$ is the number density of the system. For statistically isotropic systems, the structure factor only depends on the wavenumber $k = |{\bf k}|$. The small-$k$ scaling behavior of $S(k)$, i.e., $S({k}) \sim k^\alpha$, determines the large-$R$ asymptotic behavior of $\sigma_N^2(R)$, based on which all DHU
systems can be categorized into three classes:
$\sigma_N^2(R) \sim R^{d-1}$ for $\alpha>1$ (class I); $\sigma_N^2(R)
\sim R^{d-1}\ln(R)$ for $\alpha=1$ (class II); and $\sigma_N^2(R)
\sim R^{d-\alpha}$ for $0<\alpha<1$ (class III), where $\alpha$ is the hyperuniformity exponent  \cite{To18a}.

Besides the aforementioned DHU examples in biological systems, a wide spectrum of equilibrium \cite{To15, Ba09, jiao2022hyperuniformity} and non-equilibrium \cite{Ga02, Do05, Za11a, zachary2011hyperuniformity, yuan2021universality} many-body and material systems, in both classical \cite{Ku11, Hu12, Dr15, chen2021multihyperuniform, zhang2023approach, Ch18b} and quantum mechanical \cite{To08, Fe56, Ge19quantum, sakai2022quantum, Ru19, Sa19, Zh20, Ch21, nanotube, chen2023disordered} varieties, have been identified to possess the property of hyperuniformity. Notably, certain driven non-equilibrium systems \cite{He15, Ja15, We15, salvalaglio2020hyperuniform, nizam2021dynamic, zheng2023universal}, active-particle fluids \cite{Le19, lei2019hydrodynamics, huang2021circular, zhang2022hyperuniform, oppenheimer2022hyperuniformity}, and dynamic random organizing systems \cite{hexner2017noise, hexner2017enhanced, weijs2017mixing, wilken2022random}, which have been used as models for, e.g., self-organization of biological cells, school of fish or flock of birds, are also found to exhibit emergent hyperuniform behaviors. Novel DHU materials have been engineered that can possess superior properties compared to their crystalline counterpart, such as high-degree of isotropy and robustness against defects for transport applications \cite{Fl09, klatt2022wave, Zh16, Ch18a, torquato2021diffusion, Xu17, Le16, yu2023evolving, shi2023computational, To18b}.

Although it is not completely understood how DHU emerges in many of the aforementioned systems, the attempt to achieve certain functional optimality under constraints seems to be a crucial underlying mechanism \cite{ge2023hidden}. In the case of avian photoreceptors, the cones of the same type favor the ordered triangular-lattice arrangement that maximizes light sampling efficiency. However, this ideal arrangement is frustrated by the size polydispersity, shape irregularity and incompatible number ratios of five cone types. Therefore, the resulting DHU arrangement is one that optimizes light sampling under these constraints \cite{Ji14}. Similarly, it was found the DHU pattern of arid and semi-arid vegetation ecosystem is likely resulted from optimal utilization of water resources under constraints \cite{ge2023hidden}. These exciting discoveries naturally lead to the question: Are there any other biological systems that have developed hyperuniform organizations for optimal functionality under constraints?






\begin{figure}[ht!]
\begin{center}
$\begin{array}{c}\\
\includegraphics[width=0.485\textwidth]{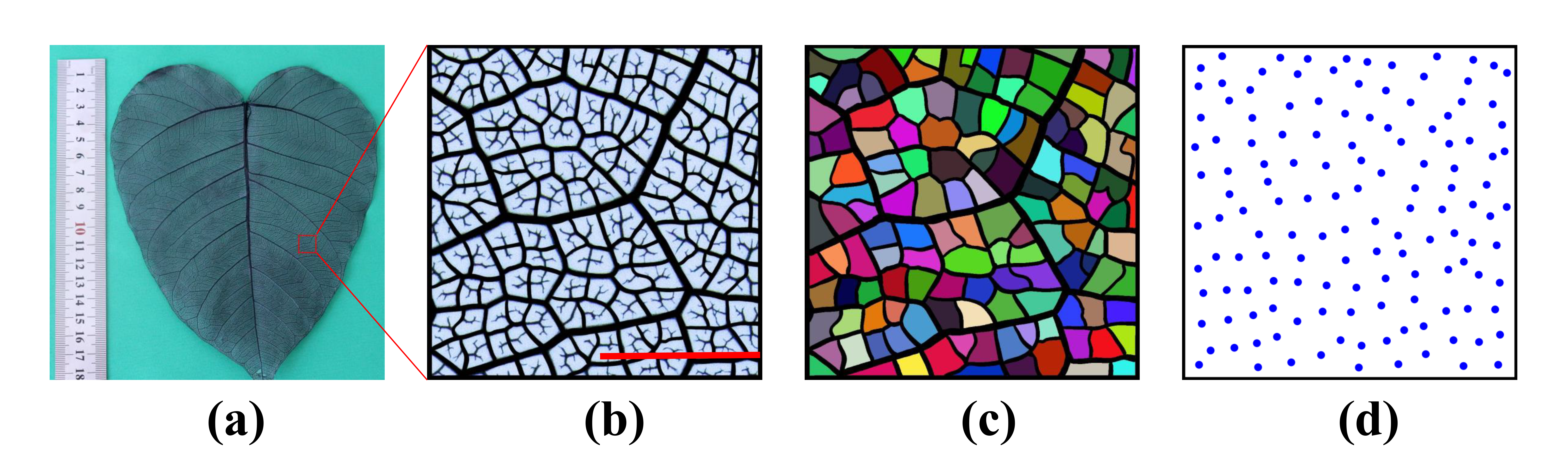} 
\end{array}$
\end{center}
\caption{Cellular structures in looped leaf vein network. (a) Optical image of air-dried ficus religiosa leaf containing a hierarchical vein network. (b) A portion of looped network formed by the secondary veins in between two primary vein branches. The scale bar is 0.5 cm. (c) Cellular structures extracted from the looped network in (b). (d) Point configuration associated with the geometrical centers of the cells in (c).} \label{fig_1}
\end{figure}


In this letter, we report the discovery of universal hyperuniform organization in looped vein networks. Leaf veins are present throughout the leaf lamina, providing mechanical support of the leaf structure \cite{scarpella2010control, biedron2018auxin}. Another key functionality of leaf veins is to transport water, minerals and nutrients to the leaf cells. These veins form a hierarchical network: On the macroscopic scale, the thick primary veins organize into a branched network, which is very efficient to mechanically support the leaf structure, see Fig. \ref{fig_1}(a). On the other hand, in between the main branches, the secondary thinner veins can form looped networks \cite{katifori2012quantifying} for efficient transport, see Fig. \ref{fig_1}(b). 

Quantitative characterizations of the venation patterns of angiosperm leafs have revealed strong local regularity \cite{ref1, ref2} and topological hierarchy \cite{katifori2012quantifying} of the vein networks, which are intimately related to their functionality. For example, it is found local vein density is strongly correlated with the transpiration rate \cite{ref1}, and the number of stomata within each areole was linearly related to the length of the looping vein contour \cite{ref2}. Katifori and Magnasco \cite{katifori2012quantifying} introduced a novel hierarchical loop decomposition framework to quantify the hierarchical topological organization of the looped venation patterns. Ronellenfitsch and Katifori \cite{ref3} showed that the development of vein networks spanning from looped to tree-like patterns corresponds to an optimal trade-off between transport efficiency, cost and robustness against defects, a property that has been observed in many disordered hyperuniform systems \cite{To18a}. 




Our detailed analysis below indicates the distribution of the geometrical centers associated with the cellular structures in the looped vein network, for all leaf species analyzed here, possesses a high degree of hyperuniformity, characterized by a vanishing structure factor in the zero-wavenumber limit, i.e., $S(k) \sim k^\alpha$ with $\alpha \approx 0.64 \pm 0.021$, providing a novel example of class III hyperuniformity in biology. Importantly, we show the DHU organization of the vein networks leads to superior transport behaviors, compared to other uncorrelated or correlated disordered organization models, which is crucial to the functionality of the veins.  



{\bf Sample preparation and image processing.} We collected and analyzed a large number of air-dried leaf samples of 6 distinct species including ficus religiosa (FR), ficus caulocarpa (FC), ficus microcarpa (FM), smilax indica (SI), populus rotundifolia (PR), and yulania denudate (YD), commonly found in middle and northern part of Asia, Europe and North America. The selected species are also representative of a larger group of plant species with looped leaf venation \cite{ref1}. We took high-resolution optical images of the samples (see Fig. \ref{fig_1}(a) for an example of ficus religiosa leaf), and used Matlab for image correction and enhancement to highlight the vein network (Fig. \ref{fig_1}(b)). We employed the OpenCV package in Python to identify the ``cells'' formed by the veins based on two criteria: (i) a cellular region is enclosed by a closed loop of veins; and (ii) a cellular region contains an open-end branched vein in the center (see SI for details). Such open-end branched structures play a similar role of capillaries, releasing the minerals and nutrients to individual leaf cells \cite{scarpella2010control}. An example of the resulting cellular structure is shown in Fig. \ref{fig_1}(c), based on which the geometrical center of each cell is obtained (with number $N$ varying from $\sim 100$ to $\sim 400$) for subsequent analysis, see Fig. \ref{fig_1}(d). The average number density of derived point configurations is $\rho \approx 125 $ cm$^{-2}$.

\begin{figure}[ht!]
$\begin{array}{c}\\
\includegraphics[width=0.45\textwidth]{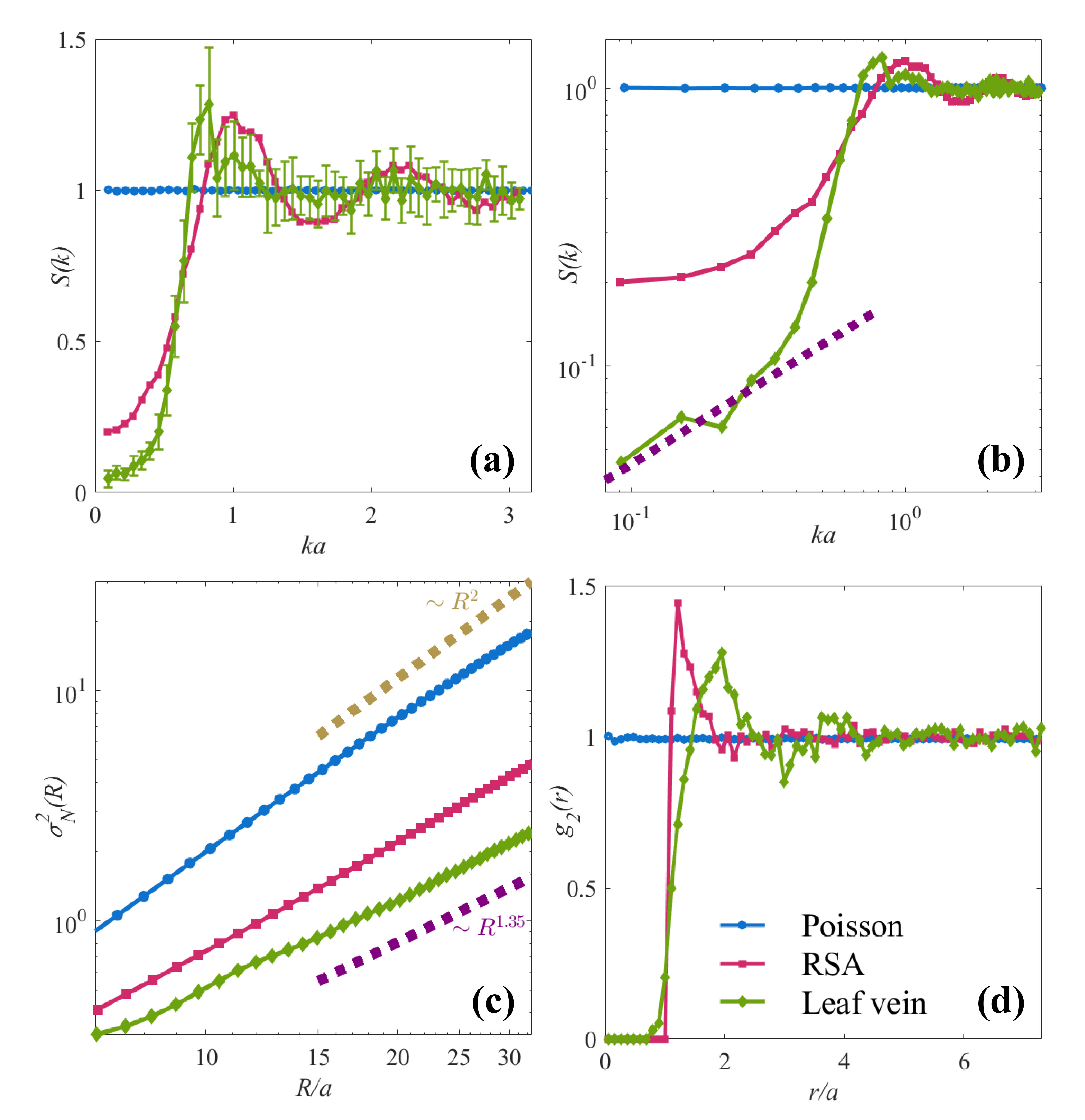} 
\end{array}$
\caption{Statistics of the point configurations with $\rho \approx 125$ cm$^{-2}$ derived from the cellular structures of the secondary vein networks of ficus religiosa leaf, including $S(k)$ (a) and (b), $\sigma_N^2(R)$ (c), as well as $g_2(r)$ (d), from 20 independent realizations (with $N \approx 200$) randomly selected from the leaf samples. The quantity $a$ is the average nearest neighbor distance in the system. The corresponding statistics of point configurations for a Poisson and random sequential addition (RSA) process with the same $\rho$ are also shown.} \label{fig_2}
\end{figure}


{\bf Hyperuniform organization of cellular structures of vein networks.} We obtain statistics of the point configurations derived from the cellular structures of the secondary vein networks, including the static structure factor $S(k)$, the number variance $\sigma_N^2(R)$, as well as the pair-correlation function $g_2(r)$, by averaging over 20 independent realizations randomly selected from the leaf samples (see SI for details). To better understand the unique hyperuniform organizations of these structures, we also consider two non-hyperuniform benchmark systems at the same number density: (i) a Poisson distribution of points and (ii) a hard-disk packing generated via random sequential addition (RSA). The former represents a totally uncorrelated disordered system, while the latter represents a correlated system through the mutual exclusion effects.   

Figure \ref{fig_2}(a) and (b) respectively shows $S(k)$ in the linear and log scale for the three systems. As expected, the uncorrelated Poisson system possesses a constant $S(k) = 1$, while the RSA system shows a strong suppression of scattering for small $k$, but is non-hyperuniform. On the other hand, the point configurations derived from the cellular structures of the leaf vein network exhibit a high degree of hyperuniformity, characterized by a vanishing $S(k)$ in the zero-$k$ limit with a hyperuniformity index \cite{To18a} $H = S(0)/S(k_{p}) \approx 10^{-3}$ (where $k_p$ is the wavevector associated the first and highest peak and $S(0) = \lim_{k\rightarrow 0}S(k)$) and a hyperuniformity exponent $\alpha \approx 0.64 \pm 0.021$ (illustrated as the dashed line), corresponding to class III hyperuniformity. To understand the boundary effects, we also carry out a system-size study and confirm that the same small-$k$ behavior and hyperuniformity exponent can be robustly obtained in systems with $N \sim 150$ to $N \sim400$ (see Fig. S2 in SI), with the latter corresponds to the largest patches of secondary veins in between the two main branches. 

Figure \ref{fig_2}(c) shows $\sigma_N^2(R)$ of the three systems in the log scale, where the Poisson and RSA systems follow the scaling $\sigma_N^2(R) \sim R^2$ and the leaf vein system possesses $\sigma_N^2(R) \sim R^\beta$, with $\beta \approx 1.35 \pm 0.017$, consistent with the estimate from $S(k)$ analysis, i.e., $ \beta = 2 - \alpha \approx 1.36$. Figure \ref{fig_2}(d) shows $g_2(r)$, characterizing the short-range correlations. Again, the Poisson system possesses a $g_2(r) = 1$, indicating its totally random nature. $g_2(r)$ of the RSA system possesses a clear exclusion region, corresponding to the diameter of the congruent circular disks, which is immediately followed by a sharp peak, resulted from the nearly contacting particle configurations. $g_2(r)$ of the vein networks is reminiscent of that of poly-disperse disk packing, characterized by a well-defined exclusion region, followed by a smooth increase and a broad peak. Similar $g_2(r)$'s were also observed for the hyperuniform distribution of the photoreceptors in avian retina \cite{Ji14}.






\begin{figure}[ht!]
$\begin{array}{c}\\
\includegraphics[width=0.49\textwidth]{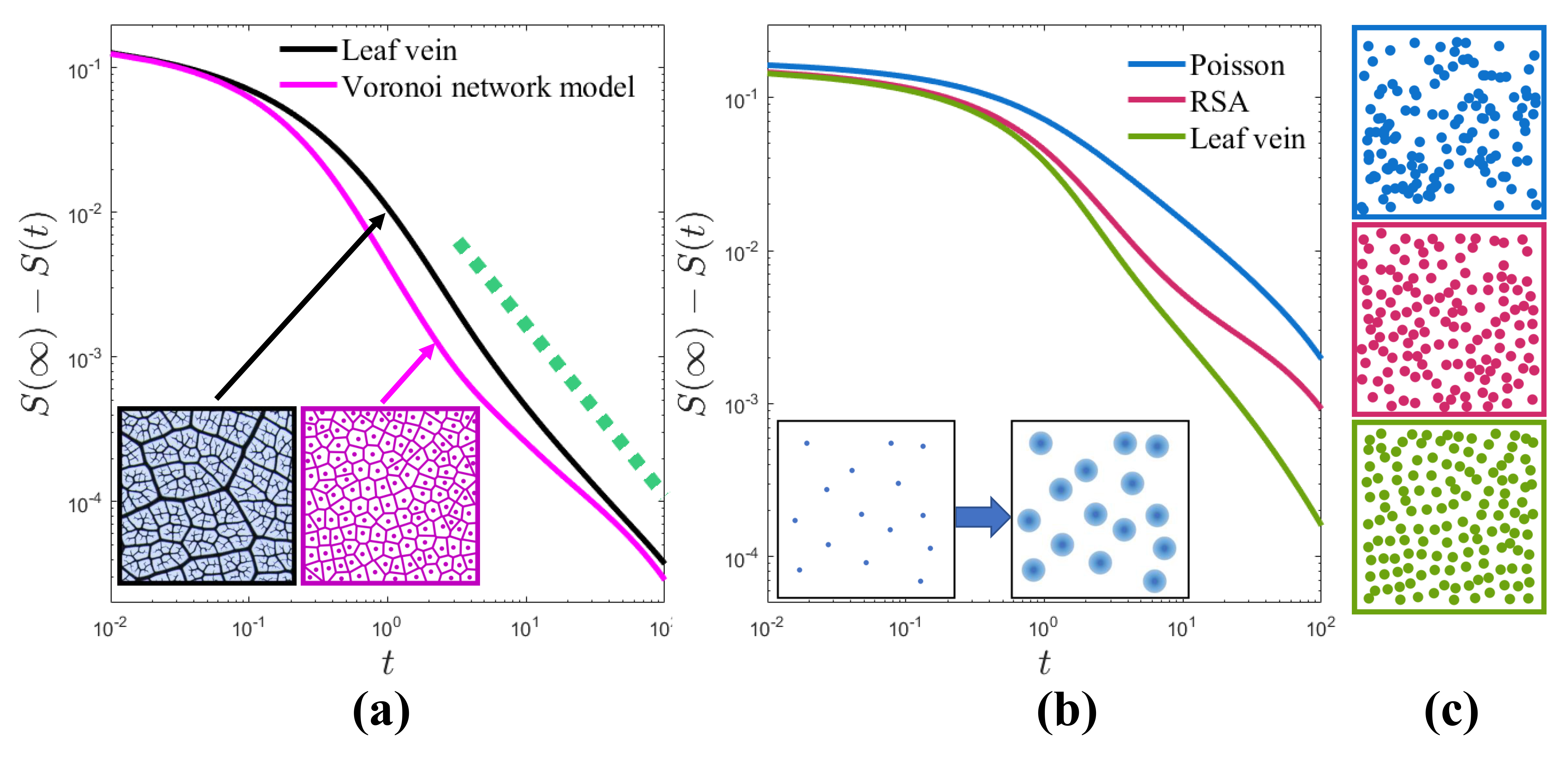} 
\end{array}$
\caption{Excess spreadability $\mathcal{S}(\infty) - \mathcal{S}(t)$ of the DHU leaf vein networks and different structural models. (a) Comparison of $\mathcal{S}(\infty) - \mathcal{S}(t)$ directly computed from vein imaging data and a model structure consisting of the Voronoi network associated with point configuration derived from the cellular vein networks, as shown in the inset. (b) Comparison of $\mathcal{S}(\infty) - \mathcal{S}(t)$ of three model structures derived from the Poisson (blue), RSA (red) and vein-network (green) point configurations by placing congruent circular disks at the points, as shown in (c). Inset of (b) illustrates the diffusion of solutes initially confined in certain sub-regions of the system.} \label{fig_3}
\end{figure}

{\bf Superior transport properties induced by hyperuniformity.} As noted above, one main function of the secondary vein network is to transport water and minerals to the leaf cells. As shown in Fig. \ref{fig_1}b, there is an open-end branched structure in the middle of each ``cellular region'' of the secondary vein network, from which the minerals can be released and diffused to the individual cells. Here we employ the {\it time-dependent diffusion spreadability} $\mathcal{S}(t)$ \cite{torquato2021diffusion} to quantify the efficiency of diffusive transport supported by the DHU structures. Specifically, a fixed amount of solutes are initially confined within a subset of regions $\mathcal{V}_2$ (e.g., within the veins), which then start diffusing into the remaining regions $\mathcal{V}_1$ at $t = 0$, assuming the diffusivity $D$ is identical for both phases (see inset of Fig. \ref{fig_3}b). $\mathcal{S}(t)$ is defined as the fraction of total solutes present in $\mathcal{V}_1$ as a function of time $t$, which is a measure of the ``spreadability'' of the solutes as a function of time. A related quantity is the excess spreadability $\mathcal{S}(\infty) - \mathcal{S}(t)$, i.e., 
\begin{equation}
\mathcal{S}(\infty) - \mathcal{S}(t) = \frac{1}{(2\pi)^2\phi_2} \int_{\mathbb{R}^2} \Tilde{\chi}_V({\bf k}) \exp[-k^2 D t]d{\bf k},
\end{equation}
where $\phi_2$ is the volume fraction of $\mathcal{V}_2$, and $\Tilde{\chi}_V({\bf k})$ is the spectral density function associated with $\mathcal{V}_2$, which is the Fourier transform of the corresponding autocovariance function $\chi_V({\bf r}) = S^{(2)}_2({\bf r}) - \phi_2^2$ and $S^{(2)}_2({\bf r})$ is the two-point correlation function of $\mathcal{V}_2$ \cite{shi2023computational}. The large-$t$ behavior of $\mathcal{S}(\infty) - \mathcal{S}(t)$ quantifies how fast the solutes can achieves the asymptotic uniform distribution in the system, reflecting the efficiency of diffusive transport.

Figure \ref{fig_3}a shows the excess spreadability $\mathcal{S}(\infty) - \mathcal{S}(t)$ of the DHU vein networks by directly computing $\Tilde{\chi}_V({\bf k})$ of the ``vein phase'' from the imaging data. Also shown is $\mathcal{S}(\infty) - \mathcal{S}(t)$ of a Voronoi network derived from the point configuration associated with the vein networks, including congruent disks centered at those points. The size of the disks is chosen such that their area is equivalent to the average area of the open-end branched structures in the areola. This simple model captures very well both short-time and long-time behavior of $\mathcal{S}(\infty) - \mathcal{S}(t)$ for the actual vein networks. Notably, the DHU vein network possesses a long-$t$ asymptotic scaling $\mathcal{S}(\infty) - \mathcal{S}(t) \sim 1/t^{\gamma}$, where $\gamma \approx 1.38$. This is to contrast the scaling $\mathcal{S}(\infty) - \mathcal{S}(t) \sim 1/t$ for a general nonhyperuniform disordered system \cite{torquato2021diffusion}. 




To further illustrate the diffusive transport efficiency of the DHU structures, we compare in Fig. \ref{fig_3}(b) $\mathcal{S}(\infty) - \mathcal{S}(t)$ of three model structures derived from the Poisson, RSA and vein-network point configurations, see Fig. \ref{fig_3}(c). For the DHU system, $\Tilde{\chi}_V({k}) = \rho \Tilde{m}^2(k) S(k)$ for the disk regions, where
$\Tilde{m}(k)$ is the Fourier transform of the shape function of the disks, and $\Tilde{m}(0)$ is a constant. Thus, the small-$k$ scaling $\Tilde{\chi}_V({k}) \sim S(k) \sim k^\alpha$ leads to $\mathcal{S}(\infty) - \mathcal{S}(t) \sim 1/t^{(1+\alpha/2)}$. The scaling analysis from the numerical data yields $\sim 1/t^{1.33}$, which is consistent with the calculated exponent $\alpha \approx 0.64$. This indicates uniform solute distribution can be achieved much faster in the DHU system than in other two disordered systems, characterized by $\mathcal{S}(\infty) - \mathcal{S}(t) \sim 1/t$. These analyses clearly indicate the superior diffusive transport efficiency of the DHU networks, compared to other disordered organizations. 

\begin{figure}[ht!]
\begin{center}
$\begin{array}{c}\\
\includegraphics[width=0.45\textwidth]{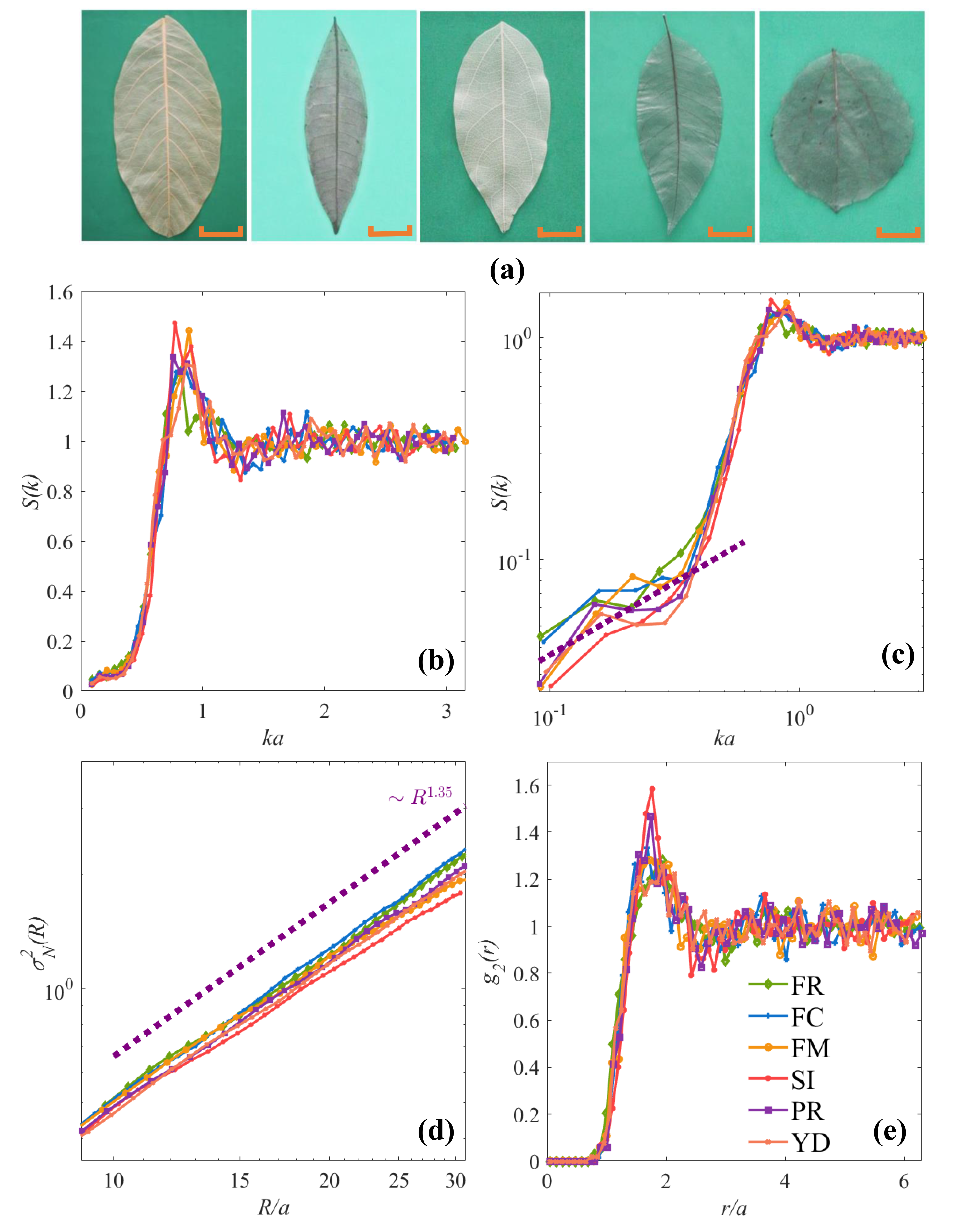} 
\end{array}$
\end{center}
\caption{Statistics of the point configurations with a rescaled density $\rho = 125 $ cm$^{-2}$ associated with distinct leaf species including ficus religiosa (FR), ficus caulocarpa (FC), ficus microcarpa (FM), smilax indica (SI), populus rotundifolia (PR), and yulania denudate (YD). The images of leaf samples are shown in (a). The scale bars are: 7.8 cm for FC; 2.6 cm for FM; 1.6 cm for SI; 3.9 cm for YD; and 2.1 com for PR. The statistics including $S(k)$ (a) and (b),  $\sigma_N^2(R)$ (c), as well as $g_2(r)$ (d) indicate universal hyperuniform organizations in the secondary vein networks of these different leaf species.} \label{fig_4}
\end{figure}


{\bf Universal hyperuniformity in looped vein networks across different leaf species.} Last but not least, we present the analyses on the leaf samples of other species including ficus caulocarpa (FC), ficus microcarpa (FM), smilax indica (SI), populus rotundifolia (PR), and yulania denudate (YD), as shown in Fig. \ref{fig_4}(a). The statistics including $S(k)$, $\sigma_N^2(R)$ and $g_2(r)$ for all six leaf species, when rescaled to the same number density, all collapse on to the corresponding universal curves, as shown in Fig. \ref{fig_4}(b)-(e). Although the small-$k$ behaviors of $S(k)$ exhibit some small discrepancy, possibly due to the finite-size effect, the estimated hyperuniformity exponent $\alpha$ for all systems agree reasonably well with one another, with the numerical values falling into the narrow interval [0.626, 0.648]. These results indicate the secondary vein networks for all six leaf species exhibit a {\it universal hyperuniform} organization on large scales, characterized by $S(k) \sim k^\alpha$ with $\alpha \approx 0.64$ which belongs to Class III hyperuniformity. In addition, the different leaf species also possess very similar local structures when properly rescaled, manifested as the collapsed pair-correlation functions.   


How such universal DHU organizations emerge during the development and regulation of the vein network? It is well established that the leaf vascular system develops in response to the flow and localization of the growth regulator auxin as continuous strands of conducting tissues arranged in repetitive spatial patterns \cite{scarpella2010control, biedron2018auxin, fujita2006origin, fujita2006pattern}. Although the auxin diffusion implies a regular length scale above which a new vein would be created, we show in SI that such local regularity and topological hierarchy do not spontaneously lead to the observed hyperuniformity, \textcolor{red}{using both an ideal geometrical model that aligns points with a narrow distance distribution along the main vein network and a more sophisticated growth model \cite{ref4} explicitly incorporating the diffusion of auxin and considers the hierarchy of vein growth.} It has been shown \cite{ref3} that hierarchically organized reticulation in venation patterns can be constructed and maintained through spatially correlated load fluctuations across length scales, offering a potential mechanism for regulating the vein networks to induce hyperuniformity.

In summary, we have discovered a universal hyperuniform organization in looped leaf vein networks across a wide spectrum of representative tree species, which does not seem to be revealed by previously reported characterizations of the venation \cite{katifori2012quantifying, ref1, ref2}, nor captured by previous models of leaf vein development \cite{ref3, fujita2006origin, fujita2006pattern, ref4}. Such DHU organization leads to superior efficiency of diffusive transport in the system. Although globally optimal transport is only achieved by periodic structures, the geometrical constraints and frustrations (e.g., curvature of primary branches and relatively short inter-branch distances) rule out the possibility to develop perfectly ordered venation patterns. In this sense, the DHU venation pattern may represent an sub-optimal organization of secondary vein networks for transport under constraints induced by the primary vein branches. 






\begin{acknowledgments}
The authors are extremely grateful to the anonymous reviewers for their constructively and valuable comments. J. T. was supported by NNSFC under Grant No. 11274200 and NSFSP  under Grant No. ZR2022MA055. J. T. thanks Dr. Cancan Xiong, Ms. Dandan Zhao, and Ms. Jingjing Liu for their help with preparing leaf samples. The codes used in this study are publicly available at https://github.com/JianxiangTian/Code.git.
\end{acknowledgments}


\end{document}